\documentclass[twocolumn,showpacs,preprintnumbers,amsmath,amssymb,floatfix]{revtex4}

\usepackage{graphicx}
\usepackage{epsfig}

\begin{document}

\title{Spin-Polarizing Cold Sodium Atoms in a Strong Magnetic Field}

\author{K.M.R. van der Stam}
\author{A. Kuijk}
\author{R. Meppelink}
\author{J.M. Vogels}
\author{P. van der Straten}
\affiliation{Atom Optics and Ultrafast Dynamics, Utrecht University, P.O. Box 80,000,
 3508 TA Utrecht, The Netherlands}

\date{\today }

\begin{abstract} 
The efficiency of evaporative cooling, which is used for the creation of a Bose Einstein condensate, depends strongly on the number of particles at the start of the evaporation. A high efficiency can be reached by filling the magneto-optical trap with a large number of atoms and subsequently transferring  these atoms to the magnetic trap as efficiently as possible. In our case (for sodium) this efficiency is limited to $1/3$, because the magnetic substates of the $F=1$ state, which is used in the trapping process, are equally populated. This limit can be overcome by spin-polarizing the sample before the transfer. For sodium atoms, however, the improvement is very small when it is done in a small magnetic field due to the large number of optical transitions in combination with the high optical density. In this article we describe spin-polarizing sodium atoms in a high magnetic field. The transfer efficiency is increased by a factor of 2. The high magnetic field makes the process also more robust against variations in the magnetic field, the laser frequency and the polarization of the laser beam.
\end{abstract}
\pacs{32.80.Bx, 32.80.Pj}

\maketitle

\section{Introduction}

\begin{figure}[ht]
	\centering
		\includegraphics[width=0.45\textwidth]{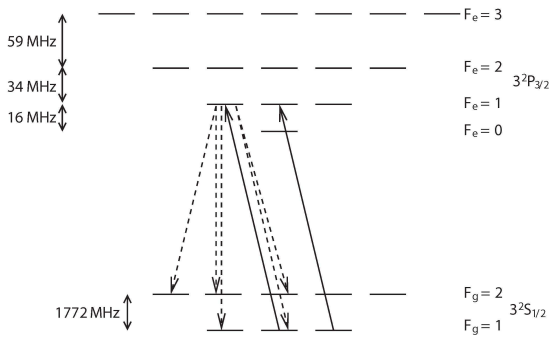}
	\caption{{The principle of spin-polarizing. The atoms in the $F_{g}=1$ ground state are optically pumped by a laser beam with $\sigma_-$ polarization (black arrows). Atoms in the $F_{e}=1$ excited state can decay back to the $F_{g}=1,2$ ground states. This is shown as an example for the $F_{e}=1, M_{g}=-1$ state (dashed arrows). Atoms in the $F_{g}=2$ ground state are pumped to the excited state by an extra laser beam with a combination of $\sigma_-$ and $\sigma_+$ polarization (not shown). }}
	\label{spinpolprincipe}
\end{figure}

The first realization of a Bose Einstein condensate (BEC) in 1995 opened up a completely new field of research \cite{Rb,Na,Li}. Currently a BEC is often used as a starting point for further research. Therefore it is of great importance to have a reliable process of creating a BEC. Creating a BEC can be divided into two main steps. First, atoms are cooled and trapped  in a magneto-optical trap (MOT) using laser cooling. Second, the sample is transferred to a magnetic trap (MT), in which evaporative cooling is applied. The latter cooling technique is based on selectively removing the most energetic atoms, followed by rethermalization of the remaining atoms due to elastic collisions. The efficiency of the evaporative cooling depends strongly on the rethermalization rate. Therefore, at the start of the evaporation the density of atoms in the magnetic trap should be as large as possible. This can be achieved by creating a MOT with a large number of atoms, and subsequently transferring those atoms to the MT as efficiently as possible. In our case, we trap sodium atoms in the $F=1$ $M=-1$ state, because only the atoms which have a decreasing potential energy with decreasing magnetic field will be trapped in the minimum of the MT (low field seekers). The transfer efficiency is limited to 1/3, because the magnetic substates of the $F=1$ state are equally occupied in the MOT. The transfer efficiency can be increased by spin-polarizing the sample before loading it into the MT. This can be done by applying a short laser pulse with the proper polarization in combination with a small magnetic field to define the different spin states with respect to the magnetic trap, after the MOT is turned off and before the MT is turned on, Fig. \ref{spinpolprincipe}. Spin-polarization is a commonly used technique in  BEC experiments, for example in the experiments with Rb \cite{Rb}, He \cite{He}, and Ne \cite{Ne}. In these experiments transfer efficiencies up to 0.70 are reached. To our knowledge in sodium BEC experiments the efficiency has never exceeded 0.35 \cite{Na2}, due to the large number of transitions lying very close to each other in frequency in combination with the high optical density. The final degree of polarization is determined by the balance between the optical pumping to the polarized state and the depolarization. The depolarization is caused by off-resonant scattering of the spin-polarization beam by atoms, which are already polarized, and the reabsoption of spontaneous emitted photons. In sodium MOTs used in BEC experiments the optical density is so high, that these are strong depolarization mechanisms. For sodium it is not possible to circumvent this problem using a larger detuning due to the large number of optical transitions, which are close to each other in frequency.\\
In this article we describe a method of spin-polarizing a high density sample of cold sodium atoms at a high magnetic field. The method suppresses the depolarization due to the previous mentioned processes strongly and makes the process also more robust against instabilities in magnetic field and laser frequency.

\section{Experimental setup}

In our BEC experiment a thermal atomic beam produced by an oven operating at 570 K is slowed longitudinally by means of Zeeman slowing. In our Zeeman slower the magnetic field decreases from approximately 1000 G at the beginning to approximately -200 G at the end of the slower. The large negative magnetic field at the end of the slower makes the extraction of the atoms from the slower less critical compared to a slower with a small positive end-field. The slowing process is driven by a laser beam with a detuning of 356 MHz below the $F_{g}=2$ to $F_{e}=3$ transition. During the passage of the zero crossing of the magnetic field atoms can be excited to the $F_{e}=2$ state, since the detuning becomes small in a small magnetic field. Atoms in this state can be lost from the slowing process due to decay to the $F_{g}=1$ ground state. Therefore, we have a second frequency (the so called repumper) which excites the atoms from the $F_{g}=1$ ground state to the $F_{e}=2$ excited state. In the case of the Zeeman slower this repump frequency is generated by adding sidebands to the laser beam with a frequency difference of 1720 MHz by using an electro-optical modulator (EOM). After the atoms have been slowed down to 30 m/s a dark-spot MOT is loaded for 6 sec. The dark-spot MOT is operated with three retro-reflected laser beams (MOT beams) with a detuning of -12 MHz with respect to the $F_{g}=2$ to $F_{e}=3$ transition, and one retro reflected beam (repump beam) with a detuning of +2.5 MHz with respect to the $F_{g}=1$ to $F_{e}=1$ transition. For the MOT the repump beam is generated with a separated laser instead of an EOM. In the dark-spot MOT configuration the repump beam has a black spot in the center of the beam. The atoms in the center of the dark-spot MOT are optically pumped in the $F_{g}=1$ ground state due to the absence of repump light \cite{darkmot}. This avoids the density limitation which is present in a bright MOT due to intra-MOT collisions \cite{intramot1,intramot2}. A typical dark-spot MOT in our experiments contains $10^{10}$ atoms at a temperature of $320$ $\mu$K at a peak density of about $3.0 \cdot 10^{11}$ cm$^{-3}$\cite{ooijen}. After the loading the magnetic field and the repump beam are first shut off and the MOT beams are shut down 1 ms later to make sure that all the atoms end up in the $F_{g}=1$ ground state. Subsequently the atoms are transferred to the magnetic trap. We use a cloverleaf trap for full 360 degree optical access with a magnetic field gradient of $100$ Gauss/cm which leads to trap frequencies in the axial and radial direction of $\nu_{z}=14.8$ Hz and $\nu_{\rho}=120.2$ Hz, respectively. Because the atoms in the MOT are equally distributed over the three magnetic sublevels of the trapped state ($F_{g}=1$), only atoms in one of the three states can be trapped after the transfer to the MT resulting in a maximum transfer efficiency of $1/3$. The efficiency can be increased by transferring the atoms to the magnetically trapped state, $F_{g}=1, M_{g}=-1$ in our case, before loading them into the MT. This spin-polarizing process can increase the number of atoms at the start of the evaporative cooling with a factor of three. Since the evaporative cooling is a non linear process, the increase of atoms in the BEC can be much larger, depending on the loss mechanism during the evaporation.

The number of atoms in our experiment can be determined either by fluorescence or absorption imaging. In this experiment we mainly use fluorescence imaging, where 20 ms after the atoms are released from the trap (MOT or MT), a 1 ms laser pulse with repump and MOT frequency is applied. The density of the cloud after an expansion time of 20 ms is small enough to neglect optical density effects. Therefore the number of atoms that are illuminated scales linearly with the measured fluorescence. We determine the transfer efficiency by taking the ratio of the number of atoms in the MOT and the MT.

A possible complication of the measurement is that not only the $F_{g}=1, M_{g}=-1$ state can be trapped in the MT, but also the $F_{g}=2, M_{g}=1,2$ states. These states are not distinguishable from the $F_{g}=1, M_{g}=-1$ state with our imaging technique. However, since we use $\sigma_-$ light and apply after spin-polarizing an extra laser pulse at the MOT ($F_{g}=2$ $\rightarrow$ $F_{e}=3$) frequency to pump the atoms to the $F_{g}=1$ state, it is highly unlikely for the atoms to end in the $F_{g}=2, M_{g}=1,2$ states.

\section{Optical transitions of Na in a magnetic field}

To determine the magnetic field and laser frequency required for the spin-polarization we have calculated the frequencies of the transitions from the $3^2S_{1/2}$ ground state to the $3^2P_{3/2}$ excited state as a function of the magnetic field for the different hyperfine levels. The results are shown in Fig. {\ref{F1} for transitions from the $F_{g}=1$ $\rightarrow$ $F_{e}=0,1,2,3$ driven by $\sigma_{-}$ and $\sigma_{+}$ light. Note that $F$ is no longer a good quantum number in a magnetic field and therefore the selection rule $\Delta F$= 0, $\pm$ 1 no longer applies. The detuning is given with respect to the $3^2S_{1/2} (F_{g}=2)$ to $3^2P_{3/2} (F_{e}=0)$ transition.

\begin{figure}[!t]
\centering
\includegraphics[width=0.45\textwidth]{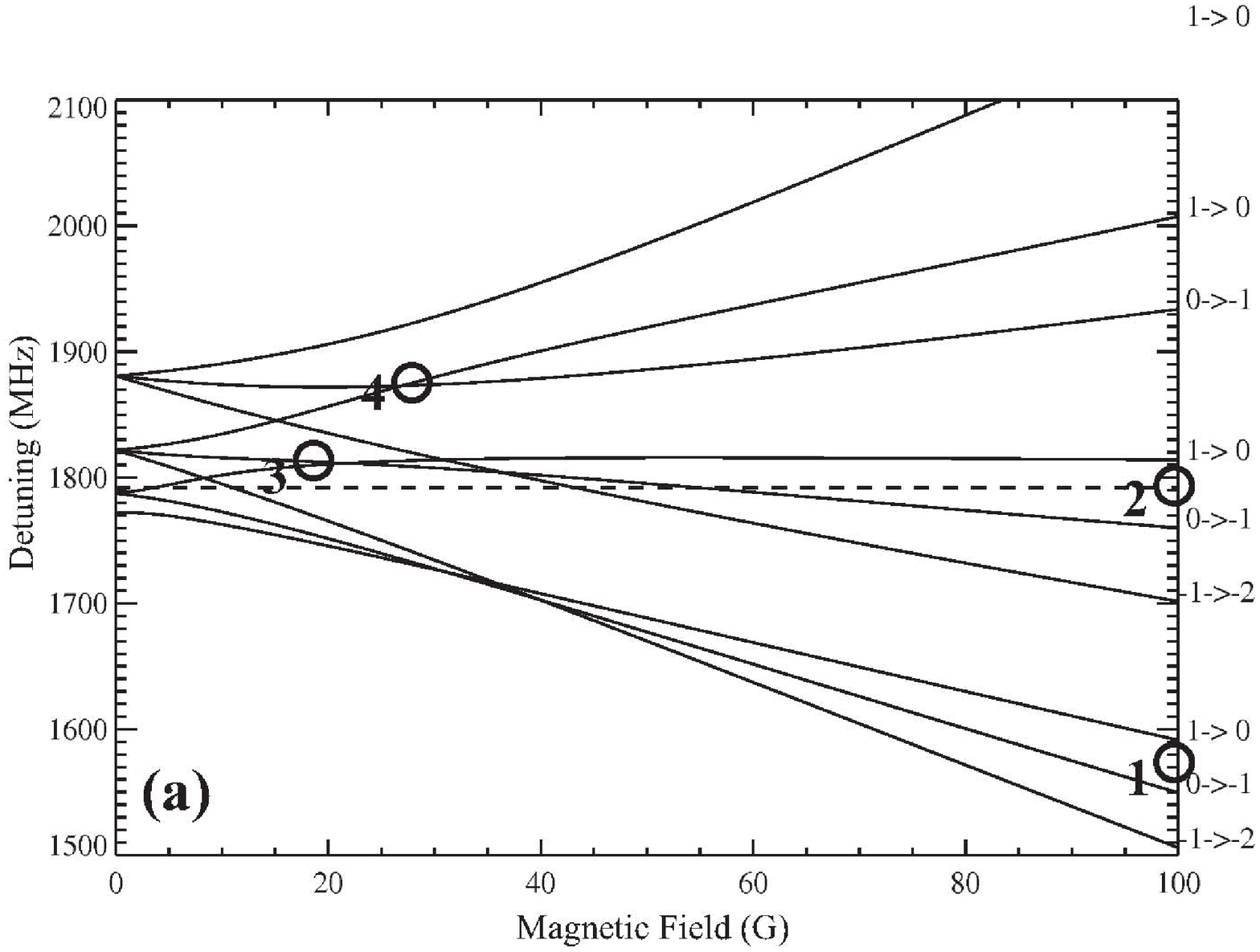} 
\includegraphics[width=0.45\textwidth]{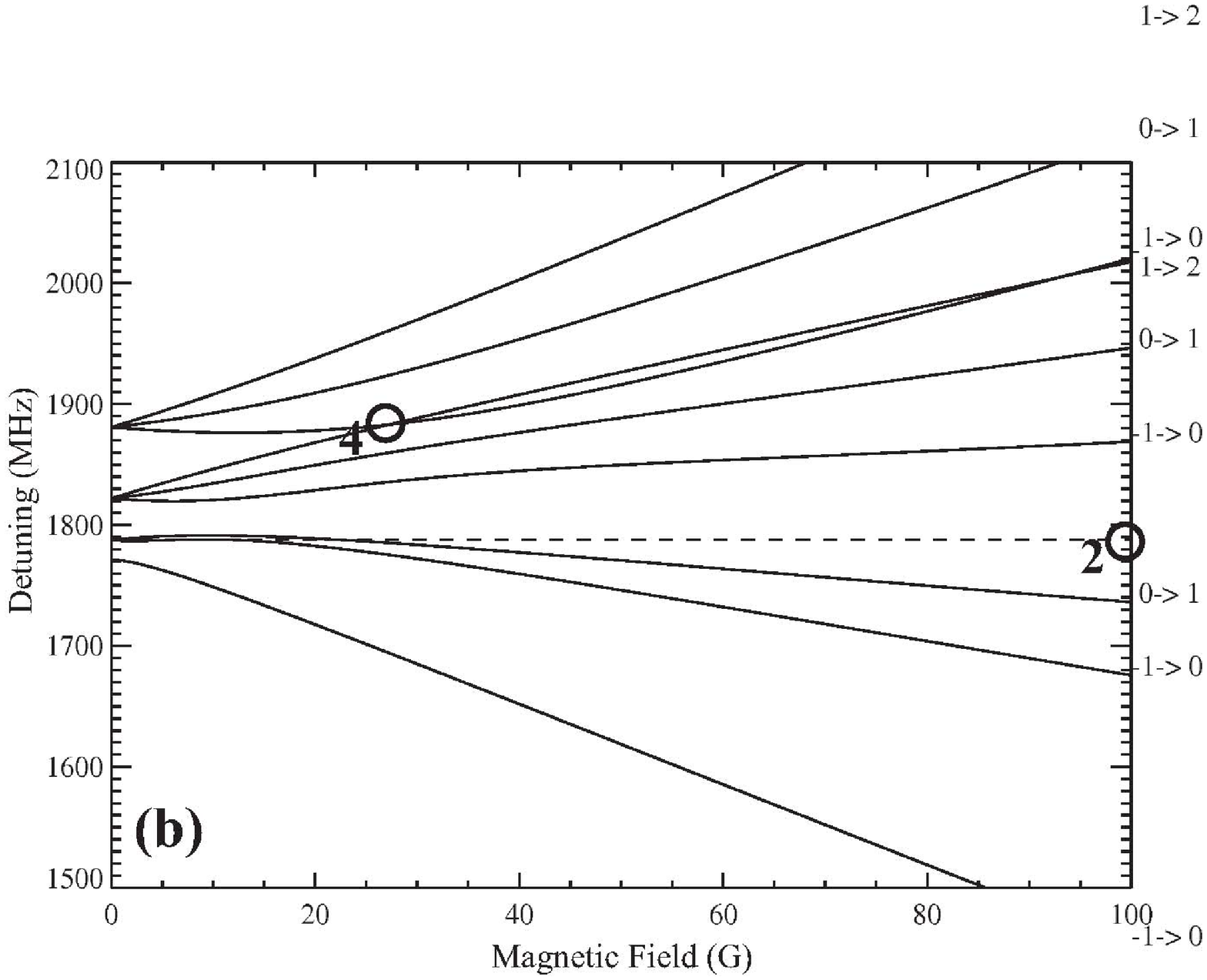} 
\caption{{The transition frequencies from the $3^2S_{1/2} (F_{g}=1)$ ground state to the $3^2P_{3/2} (F_{e}=0,1,2,3)$ excited states as a function of the magnetic field for $\sigma_{-}$ light (a) and $\sigma_{+}$ light (b). On the right side of the figure the magnetic quantum numbers of the transitions are given, M$_{g}\rightarrow$M$_{e}$. The detuning is given with respect to the $3^2S_{1/2} F_{g}=2$ to $3^2P_{3/2} F_{e}=0$ transition. The dashed line indicates the spin-polarization laser frequency. The numbers 1 to 4 in (a) indicate the possible combination of magnetic field and frequency that drive the atoms to the $F_{g}=1, M_{g}=-1$ state. In (b) the numbers 2 and 4 are the situations which are stable against fluctuations in magnetic field and frequency and can be reached in our setup (see text).}} 
\label{F1}
\end{figure}

From Fig.\ref{F1} we can determine which combinations of magnetic field and frequency result in a transfer to the sublevel, where atoms can be trapped. The first criterium is that the atoms need to be transferred to the $F_{g}=1, M_{g}=-1$ state. Therefore the transfer from $M_{g}=1$ to $M_{g}=0, -1$ state and from $M_{g}=0$ to $M_{g}=-1$ state are important. The transition from the $M_{g}=-1$ to $M_{e}=-2$ state should be suppressed because this transition is not necessary for polarizing the atoms, but can cause depolarization and heating of the cloud. The second criterium is that the situation is robust against fluctuations in magnetic field, laser frequency and polarization. The last requirement is a frequency between 1620 MHz and 1860 MHz, because detunings larger than 1860 MHz and smaller than 1620 MHz are difficult to reach in our setup.

The first requirement is met at four combinations of magnetic field and detuning shown in Fig. \ref{F1}(a), which will be discussed in this paragraph. At these combinations both transitions from $M_{g}=1$ to $M_{g}=0,$ and from $M_{g}=0$ to $M_{g}=-1$ state are close to resonance, while the $M_{g}=-1$ to $M_{e}=-2$ transition is as far detuned as possible. In case (1) a magnetic field of 100 G with a frequency of 1570 MHz is in principle possible. This detuning is, however, problematic to reach in our setup. In case (2) another suitable situation is reached at 1790 MHz, which is achievable in our setup. Case (3) is reached by decreasing the magnetic field to 20 G, while the detuning is set slightly higher at 1810 MHz. This results in a situation which also gives the two transitions we want, but it is more sensitive to fluctuations in both parameters. The last possibility (case (4)) is at a detuning of 1880 MHz, combined with 25 G magnetic field. This means that only considering the pumping process, the second and the fourth combinations are useful. Looking at the transition that can be made with $\sigma_+$ light, (see Fig. $\ref{F1}(b)$), we observe that the (25 G, 1880 MHz) combination is resonant with two transitions, while the high magnetic field (100 G) situation is not close to resonance. This means that combination 2 is much more stable against polarization fluctuations. The large energy splitting of the transitions at high magnetic field reduces also the optical density, and will for that reason be used in our work.

In the spin-polarizing process atoms are optically pumped from the $F_{g}=1$ state, because this is the occupied state in the dark-spot MOT. To prevent atoms from being lost from the spin-polarizing process due to decay to the $F_{g}=2$ we apply an extra depump frequency. This depump frequency excites the atoms from $F_{g}=2$ ground state. We use the MOT laser beams for depumping. With respect to the quantization axis of the atoms the polarization of the MOT laser beams is a combination of $\sigma_-$, $\pi$ and $\sigma_+$ polarization. In Fig. \ref{F2} the transition frequencies from the $F_{g}=2$ ground state to the different excited states as a function of the magnetic field are shown for $\sigma_-$, $\pi$ and $\sigma_+$ light, respectively.

\begin{figure}[!t]
\centering
\includegraphics[width=0.45\textwidth]{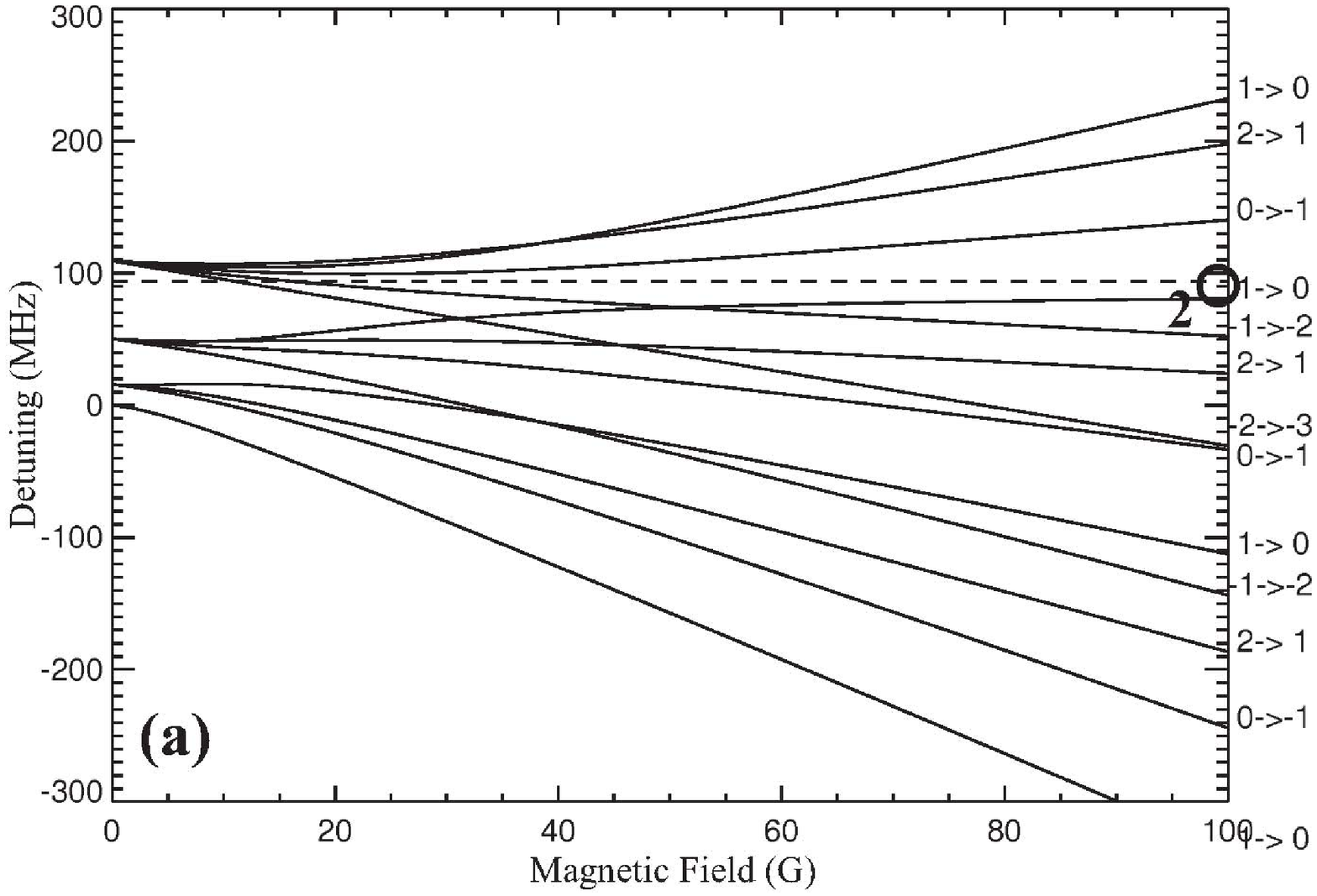}   
\includegraphics[width=0.45\textwidth]{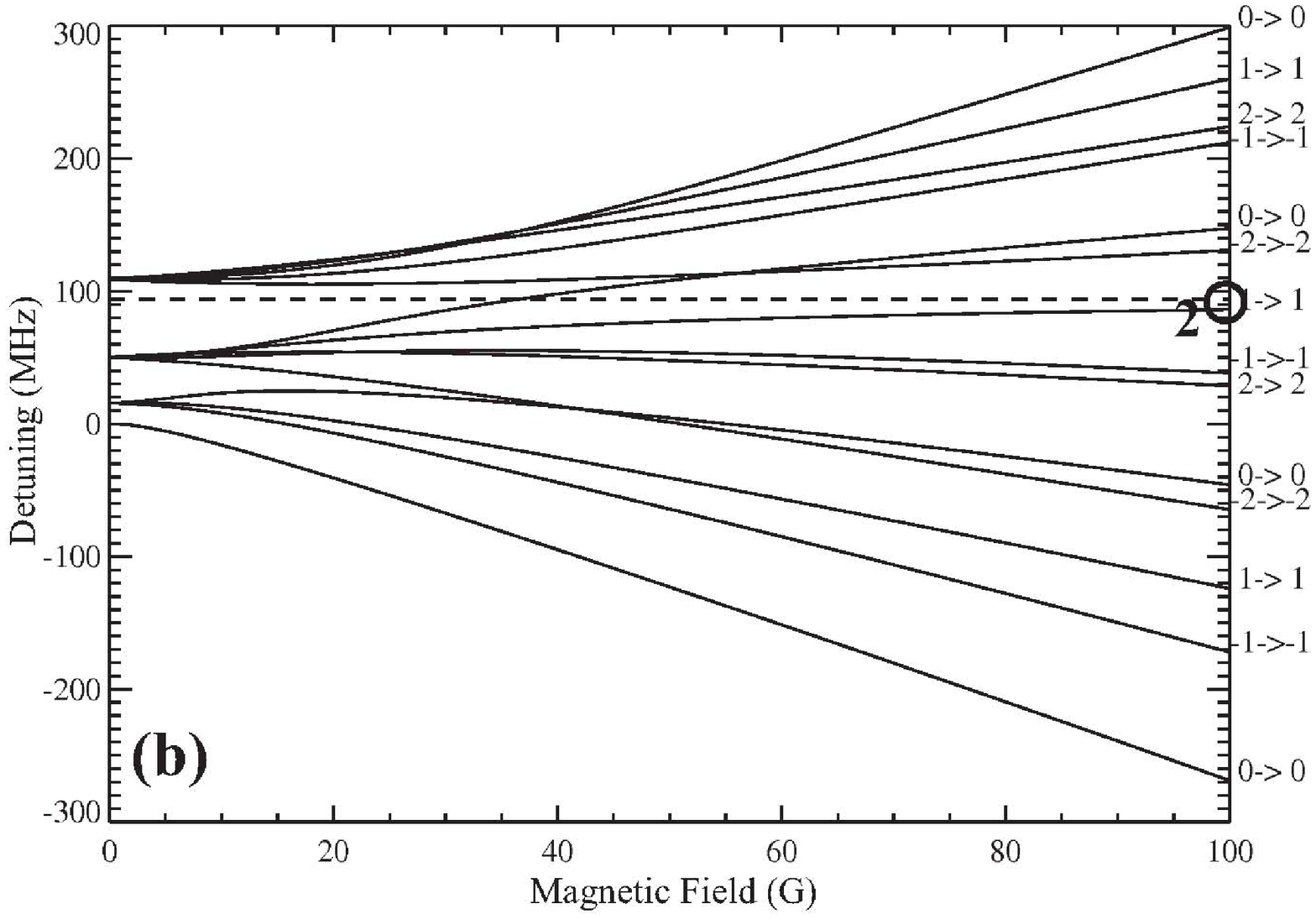}  
\includegraphics[width=0.45\textwidth]{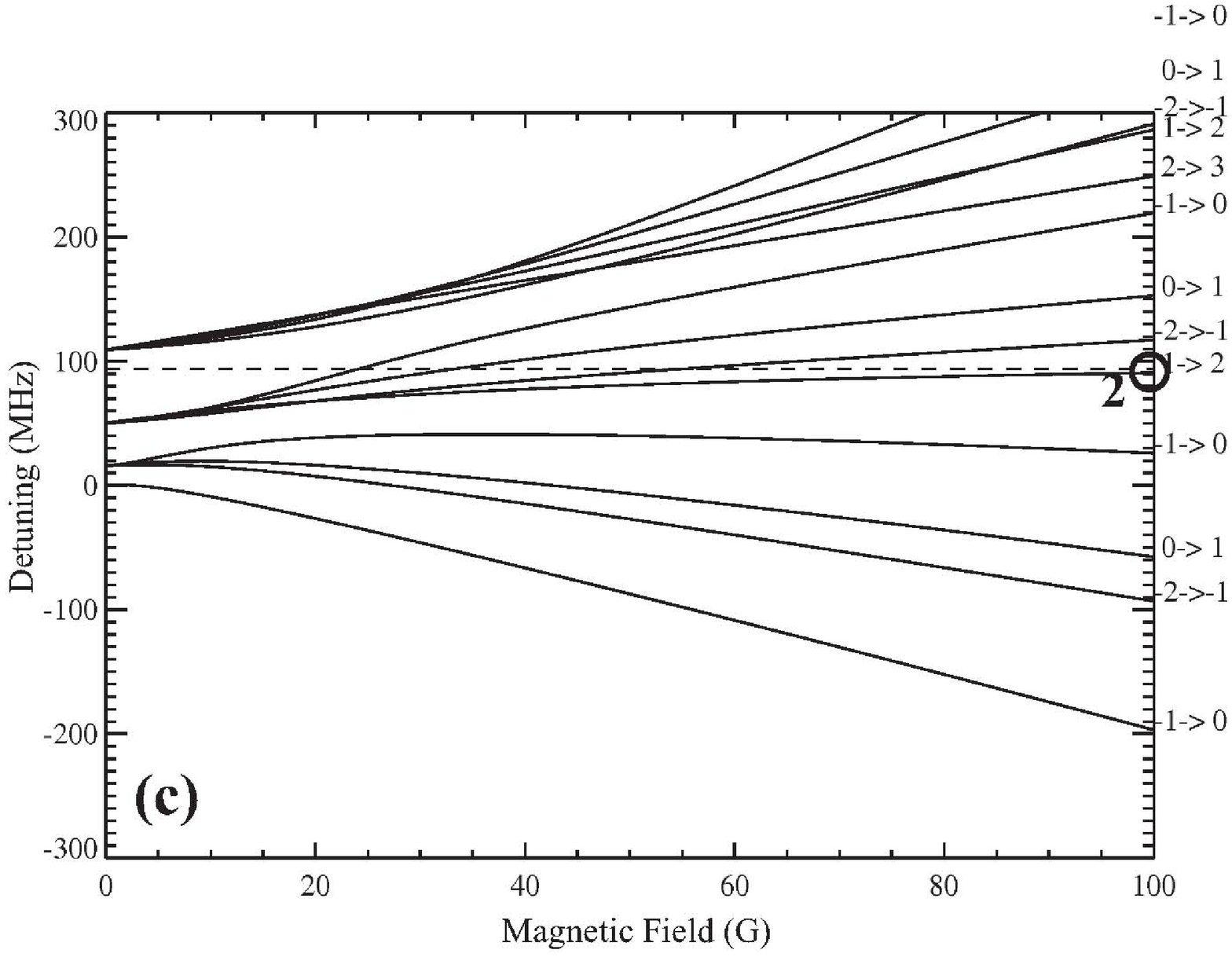} 
\caption{{The transition frequencies from the $3^2S_{1/2} (F_{g}=2)$  ground state to the $3^2P_{3/2} (F_{e}=0,1,2,3)$ excited states as a function of the magnetic field for $\sigma_-$(a), $\pi$(b), and $\sigma_+$(c) light. To the right side of the figure the magnetic quantum numbers M$_{g}\rightarrow$M$_{e}$ of the transition are given.  The detuning is given with respect to the $3^2S_{1/2} (F_{g}=2)$ to $3^2P_{3/2} (F_{e}=0)$ transition. The dashed line indicates the used laser frequency. Number 2 gives the combination of magnetic field and laser frequency we use for the depumping during spin-polarizing.}}
\label{F2}
\end{figure}

Figure \ref{F2} shows that at the chosen combination (100 G and 1790 MHz) the depump frequency is nearly resonant with the $F_{g}=2, M_{g}=1$ to $F_{e}=2, M_{e}=0,1,2$ transition. However, the other magnetic sublevels of the ground state are much further detuned. When atoms pass another $F_{g}=2$ states during the optical pumping process it takes more time to complete the spin-polarization. The scatter rate for an off-resonant transition is given by  
\begin{equation}
	\gamma_s=\frac{s_0\gamma/2}{1+s_0+(2\delta/\gamma)^2},
\label{spinpoleq1}
\end{equation}
where $s_0$ is the saturation  parameter, $\delta$ is the detuning from resonance and $\gamma$ is the line width of the used transition.
For example, the $F_{g}=2, M_{g}=0$ has a detuning of 113 MHz with respect to the depump frequency. The scatter rate ($\gamma_s$) from this state decreases by more than a factor of 500 with respect to the on-resonant scattering.

To simulate the optical pumping process, we have to solve the Optical Bloch equations (OBE) for the density matrix of a multi-level atom in a magnetic field. In case the optical pumping process is long compared to the lifetime for spontaneous decay, we can adiabatically eliminate the optical coherences and obtain generalized equations for the evolution of the density matrices $\sigma_g$ of the ground state and $\sigma_e$ of the excited state~\cite{nienhuis}. To simplify the treatment we choose the quantization axis along the propagation direction of the pump laser, in which case only the diagonal elements of the density matrix play a role. The evolution equations of the density matrix thus reduce to rate equations for the populations of the ground and excited state. The effect of the depumping process is taken into account by assuming that the depumper beam induces all ($\sigma_+$, $\pi$ and $\sigma_-$) possible transitions in the atom with equal strength. Finally, we eliminate the population of excited states. This is valid, since the duration of optical pumping process is much longer than the excited state lifetime. This way the largest rate in the rate equations is reduced from the spontaneous decay rate to the optical pump rate, which in case of large detuning can be much smaller. This allows us to increase the time step in the numerical integration of the rate equation considerably and thus decrease the computation time.\\
The optical pumping process can now be described as the transfer of the atom from the ground to the excited state by the absorption of a photon, followed by a branching of the atom from one excited state to different ground states by spontaneous emission. The transfer of an atom from one ground state $i$ to another ground state $j$ is than determined by an optical pump rate $\gamma_p$, which in zero magnetic field is proportional to the Clebsch-Gordan coefficient $C_{eg}^q$ of the polarization $q=0,\pm1$ of the pump beam, the detuning $\delta$ of the transition and the branching ratio $B_{je}$ of the spontaneous emission process:
\begin{equation}
\gamma_{p,ij} = \sum_e \frac{s_0 B_{je} C_{ei}^q{}^2 \Gamma/2}{1+s_0 C_{ei}^q{}^2 +(2 \delta/\Gamma)^2},
\label{spinpoleq2}
\end{equation}
where the summation runs of all possible excited states $e$. Here $s_0=I/I_{\rm sat}$ is the saturation parameter for the strongest transition and $ \delta = \omega -\omega_0$. In case of a non-zero magnetic field, all these three parameters are modified. First, the magnetic field changes the energies of ground and excited states and therefore the transition frequencies $\omega_0$. Second, since the states at non-zero field become a linear combination of the states at zero field, the coupling between ground and excited states for a given polarization is modified and this effects both the coupling coefficient $C_{ei}^q$ and the branching ratio $B_{je}$. All these effects are taken into account in our simulation.

\section{spin-polarizing in a high magnetic field}

The process of spin-polarizing in a high magnetic field as indicated in the previous section is done at a field of 100 G combined with a detuning which equals 1790 MHz. The magnetic field is created by a combination of the field produced by the pinch and the bias coils of our cloverleaf magnetic trap. This field can be tuned from 0 G to 125 G with an accuracy of 1 G. The frequency of the spin-polarizing laser can be tuned by changing the RF frequency that drives the AOM and can be changed from 1790 MHz to 1850 MHz with an accuracy of 3 MHz.

To determine the efficiency of the spin-polarizing process we measure the transfer from the MOT to the MT. This is done by loading the atoms in the dark-spot MOT for 6 seconds. After the loading the MOT laser beams and the MOT magnetic field are turned off and subsequently the sample is spin-polarized by switching on the magnetic field, the spin-polarizing laser beam and the MOT laser beams (for depumping) for a certain period $\tau$. After switching the magnetic field and the spin-polarizing beams off, we apply an extra depump period ($\tau_{depump}$) to make sure all the atoms end up in the $F_{g}=1$ ground state. After 6 seconds of trapping the atoms in the MT, the MT is turned off and subsequently fluorescence imaging is applied, as described in section 2. The transfer efficiency is determined by taking the ratio between the fluorescence images of the MOT and the MT. The efficiency depends on the polarization and the pulse length ($\tau$) of the spin-polarizing beam, the extra repump time ($\tau_{depump}$), the magnetic field, the frequency of the spin-polarizing beam ($\omega$) and the frequency of the depump beam ($\omega_{depump}$). All data is taken, unless otherwise stated, with a polarization beam intensity of $5$ mW/cm$^2$, a repump beam intensity of $20$ mW/cm$^2$, $\tau$ = 1 ms, $\tau_{repump}$ = 1 ms and a magnetic field of 100 G.

In Fig. \ref{polarization} the efficiency of the transfer is shown as a function of the polarization of the  spin-polarizing beam.
\begin{figure}[htb]
\centering
\includegraphics[width=0.45\textwidth]{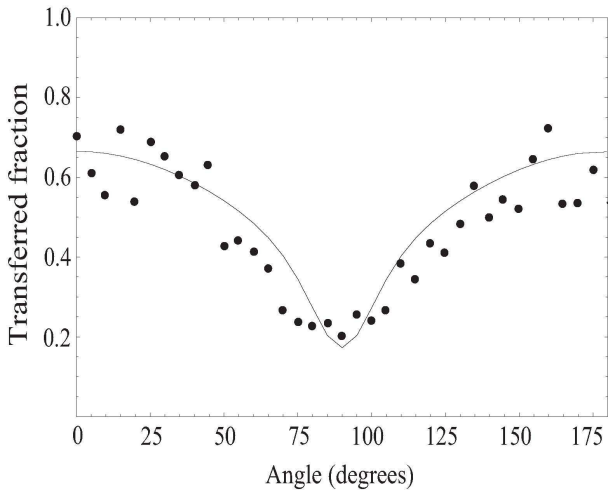}
\caption{{The transfer efficiency from the MOT to the MT as a function of the angle between the linear polarization of the laser beam and the slow axis of a quarter wave plate. Here 0, 45 and 90 degrees correspond with $\sigma_-$, $\pi$ and $\sigma_+$, respectively. The dots are the measurements and the solid line is the simulation.}}
\label{polarization}
\end{figure}
The maximum measured transfer fraction is 0.75, which implies an increase with a factor of 2.7, compared to 0.28 without spin-polarizing. That the fraction of 0.28 is lower than 1/3 is due to imperfections in the geometrical overlap between the two traps. In day to day measurements we obtain an increase with a factor of 2. Furthermore, we can see that the simulation agrees with the measurements satisfactory and we will use the results of the simulation to explain other data.

The used pulse length of 1 ms is long compared to the natural lifetime of sodium of 16 ns. In 1 ms an atom can scatter on resonance more than $6\cdot10^4$ photons, which is much more than the number of photons it takes to spin-polarize the atom, which is of the order of 10. We have measured the transfer efficiency of the atoms from the MOT to the MT as a function of the pulse length (Fig. \ref{pulsduur}). 
\begin{figure}[htb]
\centering
\includegraphics[width=0.45\textwidth]{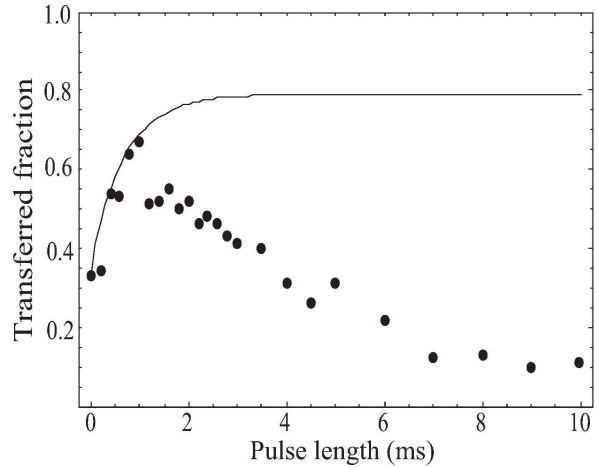}
\caption{{The transfer efficiency of atoms from the MOT to the MT as a function of the pulse length of the spin-polarizing pulse.}}
\label{pulsduur}
\end{figure}
We observe an increase of the transfer efficiency until a pulse duration of 1 ms and a decrease of the transfer towards longer pulses. To understand this behavior qualitatively we have to take the decrease of the scatter rate by a factor of 500 due to the detuning of the depump transitions and a decrease by a factor of 15 due to the detuning of the spin-polarization transition into account. We find that on average only 8.4 photons per atom are scattered during the pumping process. It becomes problematic that the $F_{g}=1, M_{g}=-1$ state is not completely dark. Although we set the frequency such that the $F_{g}=1, M_{g}=-1$ to $F_{e}=3, M_{g}=-2$ transition is off resonance, the scatter rate is still more than 8000 s$^{-1}$. Those emitted photons will be reabsorped by the optically thick sample and can cause a depolarisation of the sample. Our simulation fails to predict this, because we do not take the reabsorption into account. Why at longer pulse durations ($>$1 ms) the transfer fraction decreases is not clear to us, but we have not investigated this in more detail.

In Fig. \ref{magneticfield} the results of the measurement are shown, in which the transfer of the atoms is determined as function of the magnetic field.  
\begin{figure}[htb]
\centering
\includegraphics[width=0.45\textwidth]{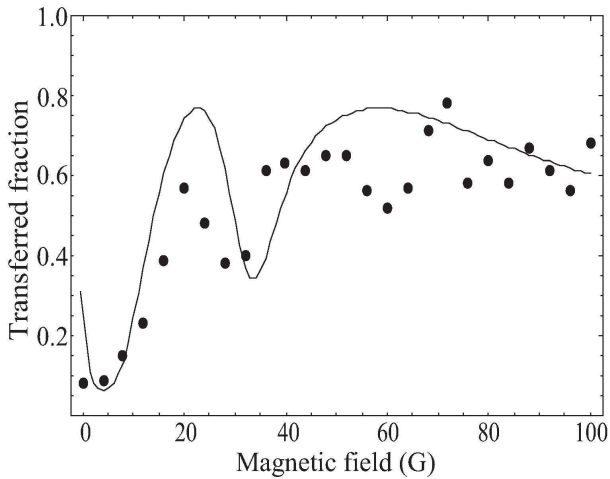}
\caption{{The transfer efficiency of atoms from the MOT to the MT as a function of the magnetic field applied during spin-polarization.}}
\label{magneticfield}
\end{figure}
These results show that the transfer is large and nearly constant at magnetic fields in excess of 40 G. Therefore the process is insensitive for fluctuations in the field strength in this range. 
The minima at 10 G and 35 G can be understood by looking at Fig. \ref{F1}a. The dotted line (the used laser frequency) crosses the $F=1, M_{g}=-1$ to $F=2, M_{g}=-2$ transition at 10 G and the $F=1, M_{g}=-1$ to $F=3, M_{g}=-2$ transition at 35 G. At these magnetic fields the atoms in the $F=1, M_{g}=-1$ state scatter more than $10^4$ photons in 1 ms, which results in complete depolarization of the sample.

At fields below 5 Gauss, which are in generally used for spin-polarization, the measurements show a small transfer efficiency. This is due to the large detuning of the depump frequency (30 MHz), causing most of the atoms to end up in the $F_{g}=2$ state. This can be circumvented by using an extra depump laser beam at the resonance frequency instead of using the MOT laser beams. However, even with an extra depump beam the spin-polarizing at fields below 5 Gauss is problematic according to our simulation because the detuning of the $F_{g}=1, M_{g}=-1$ $\rightarrow$ $F_{e}=2, M_{e}=-2$ transition is too small to suppress this transition significantly. Furthermore, since the sample is optically thick the absorption of the spontaneously emitted photons leads to depolarization of the sample \cite{Na2}}. 

\begin{figure}[htb]
\centering
\includegraphics[width=0.45\textwidth]{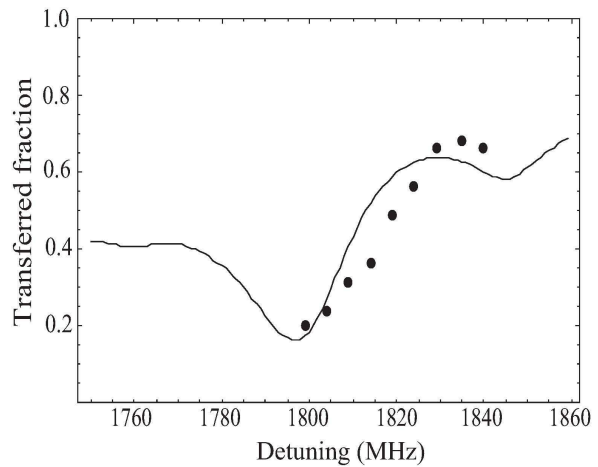}
\caption{{The transfer efficiency of atoms from the MOT to the MT as a function of the spin-polarization frequency.}}
\label{frequency}
\end{figure}

The results of the measurement of the frequency dependence of the spin-polarization are shown in Fig. \ref{frequency}. This measurement is done at a magnetic field strength of 60 G instead of 100 G. The results can be explained using Fig. \ref{F1}. When  we increase the detuning at a constant magnetic field of 60 G, the essential transitions $F_{g}=1, M_{g}=0$ $\rightarrow$ $F_{e}=2, M_{e}=-1$ and $F_{g}=1 M_{g}=1$ $\rightarrow$ $F_{e}=2, M_{e}=0$  are reached at 1800 MHz. Both the measurements and the simulation show indeed an increases at 1800 MHz.

The dependence on the depump time has also been measured. It turns out that a $\tau_{depump}$ of 1 ms is optimal. This long period (1 ms) is due to the large detuning of the depump beam. This means that during spin-polarizing a significant fraction of the atoms occupies the $F_{g}=2$ state. A frequency change of $\pm$ 10 MHz with respect to $\omega_{depump}$ does not influence the transfer efficiency significantly, but due to the detuning the necessary pulse length changes.

\begin{figure}[htb]
\centering
\includegraphics[width=0.45\textwidth]{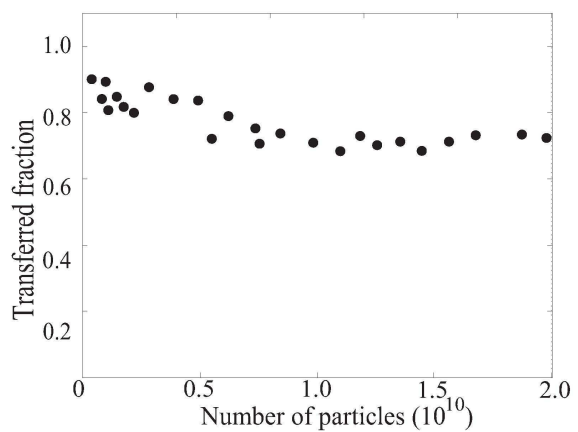}
\caption{The transfer efficiency of atoms from the MOT to the MT as a function of the number of particles in the MOT. Under normal conditions the number of atoms in the MOT equals 1$\cdot$10$^{10}$.}
\label{transfervsmot}
\end{figure}

The influence of the reabsorption of the emitted photons by the optically thick sample is studied by measuring the transfer efficiency as a function of the number of atoms in the MOT. In Fig. \ref{transfervsmot} we see a small increase in the transfer efficiency when the number of particles is decreased. As the density scales with the number of particles \cite{darkmot}, spin-polarizing works only slightly less when the optical density is increased by more than one order of magnitude. This shows that reabsorption is strongly suppressed in our experiment by using a high magnetic field. This agrees qualitatively with the theoretical work of Tupa and Anderson \cite{Tupa}, where they study the effect of radiation trapping on the polarization of a cloud of thermal atoms. However, their results can not be applied to our work, since in their case the reabsorption takes place in a Doppler-broadened sample.

\begin{figure}[htb]
\centering
\includegraphics[width=0.45\textwidth]{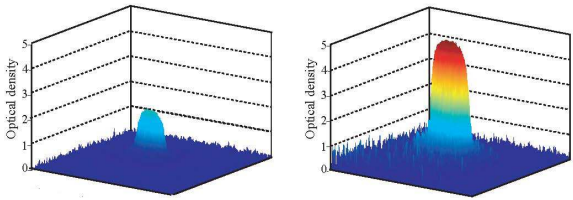}
\caption{{A three-dimensional presentation of the effect of spin-polarization after evaporative cooling using absorption imaging after an expansion time of the cloud of 40 ms. The left picture shows the optical density of a BEC produced without using spin-polarization, the right picture shows the same measurement with spin-polarization. The left condensate contains 1 $\cdot$10 $^{6}$ atoms. spin-polarizing the sample enhances the number of atoms to 9 $\cdot$10 $^{6}$(right picture), which implies a gain of a factor of 9 in the condensate.}}
\label{BEC}
\end{figure}

Finally, we studied the effect of the gain of a factor of 2 in transfer efficiency on the evaporative cooling of the trapped atoms in order to reach Bose Einstein condensation. 
In Fig. \ref{BEC} an image of a BEC without and with the use of spin-polarization is shown. Since evaporative cooling is a non-linear process this gain factor of 2 in the number of particles at the beginning of the cooling leads in Fig. \ref{BEC} to an increase of a factor of 9 in the number of particles in the BEC.
The latter gain depends strongly on the number of particles at the start of the evaporation. After we improved our initial conditions, we start with a higher number of particles in the MT. In this case the loss of atoms due to three-body collisions during the evaporation limits the increase in the BEC. Furthermore, the increase in efficiency of the evaporation process is less, if the initial number of particles increases. In the situation, that without spin-polarizing the sample results in a BEC containing 6$\cdot$10$^{6}$ atoms, the gain becomes a factor of 3. This results in a BEC of 1.8$\cdot$10$^{7}$ atoms.

\section{Conclusion}

In this article we have described spin-polarizing of a sample of cold sodium atoms in a high magnetic field. This way of spin-polarizing circumvents the problem at low magnetic fields with the large number of optical transitions in combination with the high optical density. Furthermore, this method is more robust against variations in the applied magnetic field as well as the used frequency. We have increased the transfer efficiency of the atoms from the MOT to the MT by a factor of maximum 2.7. The day to day gain is a factor of 2. This increase leads to a gain of a factor of 3--9 in the condensate. 

\section{Acknowledgments}

We acknowlegde stimulating discussions with E.D. van Ooijen. We thank H.G.M. Heideman for helpful discussions during the preparation of this manuscript.


\begin{thebibliography}{99}

\bibitem{Rb} M.H. Anderson, J.R. Ensher, M.R. Matthews, C.E.
Wieman, and E.A. Cornell, Science {\bf 269}, 198 (1995)

\bibitem{Na} K.B. Davis, M.-O. Mewes, M.R. Andrews, N.J. van
Druten, D.S. Durfee, D.M. Kurn, and W. Ketterle, Phys. Rev. Lett.
{\bf 75}, 3969 (1995)

\bibitem{Li} C.C. Bradley, C.A. Sackett, J.J. Tollett, and R.G.
Hulet, Phys. Rev. Lett. {\bf 75}, 1687 (1995); C.C. Bradley, C.A.
Sackett, and R.G. Hulet, Phys. Rev. Lett. {\bf 78}, 985 (1997)

\bibitem{He} F. Pereira Dos Santos, J. Leonard, J. Wang, C.J. Barrelet, 
F. Perales, E. Rasel, C.S. Unnikrishnan, M. Leduc, and C. Cohen-Tannoudji, 
Phys. Rev. Lett. {\bf 86}, 3459 (2001)

\bibitem{Ne} S.J.M. Kuppens, J.G.C. Tempelaars, V.P. Mogendorff, B.J. Claessens, 
H.C.W. Beijerinck, and E.J.D. Vredenbregt, Phys. Rev. A {\bf 65}, 023410 (2002)

\bibitem{Na2} Z. Hadzibabic, S. Gupta, C.A. Stan, C.H. Schunck, M.W. Zwierlein, K. Dieckmann, and 
W. Ketterle, Phys. Rev. Lett {\bf 91}, 160401 (2003)

\bibitem{darkmot} W. Ketterle, K.B. Davis, M.A. Joffe, A. Martin, and D.E. Pritchard, 
Phys. Rev. Lett. {\bf70}, 2253 (1993)

\bibitem{intramot1} M. Prentiss, A. Cable, J.E. Bjorkholm, S. Chu, E.L. Raab, and 
D.E. Pritchard, Opt. Lett. {\bf13}, 452 (1988)

\bibitem{intramot2} T. Walker, D. Sesko, and C. Wieman, 
Phys. Rev. Lett. {\bf64}, 408 (1990)

\bibitem{ooijen} E.D. van Ooijen, \textit{Realization and Illumination of Bose-condensed Sodium Atoms}, 
Ph. D. thesis (Utrecht University 2005)

\bibitem{nienhuis} G. Nienhuis, J. de Phys. C{\bf 1}, 97 (1985).

\bibitem{Tupa} D. Tupa, and L.W. Anderson, Phys. Rev. A{\bf36}, 2142 (1987)

\end{thebibliography}
\end{document}